\documentclass[conference]{IEEEtran}
\ifCLASSOPTIONcompsoc
  \usepackage[nocompress]{cite}
\else
  \usepackage{cite}
\fi
\ifCLASSINFOpdf
  \usepackage[pdftex]{graphicx}
\else
\fi
\ifCLASSOPTIONcompsoc
 \usepackage[caption=false,font=footnotesize,labelfont=sf,textfont=sf]{subfig}
\else
 \usepackage[caption=false,font=footnotesize]{subfig}
\fi
\usepackage{xcolor}
\usepackage{amsmath}
\usepackage{amssymb}
\usepackage{booktabs}
\usepackage{makecell}
\usepackage{pifont}
\usepackage{longtable}
\usepackage{array}
\usepackage{multirow}
\usepackage{xspace}
\usepackage{wasysym}

\newcommand{\yes}[1]{\CIRCLE}
\newcommand{\no}[1]{\Circle}
\newcommand{\impt}[1]{\bigstar}
\usepackage{listings}
\newcommand{\code}[1]{\lstinline[basicstyle=\ttfamily]{#1}}
\newcommand{\sys}{\textsc{RNGGuard}\xspace}

\newcommand{\mlparagraph}[1]{\vspace{.3em}\noindent\textbf{#1}}

\usepackage{url}
\begin{document}
\title{One RNG to Rule Them All: How Randomness Becomes an Attack Vector in Machine Learning}
%% add authors
\author{\IEEEauthorblockN{Kotekar Annapoorna Prabhu, Andrew Gan, Zahra Ghodsi}
\IEEEauthorblockA{\textit{Elmore Family School of Electrical and Computer Engineering, 
Purdue University}\\
\{prabhu21, gan35, zahra\}@purdue.edu}
}

\maketitle

\begin{abstract}
    Machine learning relies on randomness as a fundamental component in various steps such as data sampling, data augmentation, weight initialization, and optimization. Most machine learning frameworks use pseudorandom number generators as the source of randomness. However, variations in design choices and implementations across different frameworks, software dependencies, and hardware backends along with the lack of statistical validation can lead to previously unexplored attack vectors on machine learning systems. Such attacks on randomness sources can be extremely covert, and have a history of exploitation in real-world systems.
    In this work, we examine the role of randomness in the machine learning development pipeline \emph{from an adversarial point of view}, and analyze the implementations of PRNGs in major machine learning frameworks.
    
    We present \sys to help machine learning engineers secure their systems with low effort. \sys statically analyzes a target library's source code and identifies instances of random functions and modules that use them. At runtime, \sys enforces secure execution of random functions by replacing insecure function calls with \sys's implementations that meet security specifications. Our evaluations show that \sys presents a practical approach to  close existing gaps in securing randomness sources in machine learning systems.
\end{abstract}


% \fi
\IEEEpeerreviewmaketitle

%%%%%%%%%%%%%%%% INITIAL SUBMISSION
% \input{Sections/intro}
% \input{Sections/background}
% \input{Sections/threatmodel}
% \input{Sections/methodology}
% \input{Sections/patcher}
% \input{Sections/eval}

% \input{Sections/relwork}
% \input{Sections/conclusion}
%\input{Sections/discussionconclusion}

%%%%%%%%%%%%%%%%% CAMERA READY
\section{Introduction}
Machine Learning (ML) systems increasingly rely on randomness across various stages, such as data sampling, augmentation, weight initialization, optimization, as well as defenses against privacy or adversarial attacks such as differential privacy~\cite{dp} or robustness certification~\cite{RSmoothingOG} Randomness is fundamental for ensuring model generalization, breaking symmetry in neural networks, and improving the diversity of model outputs. However, the role of randomness in ML has largely been overlooked from a security perspective. In most ML frameworks, Pseudorandom Number Generators (PRNGs) are the primary source of randomness.  

While these PRNGs are designed to produce numbers that appear random, their implementation varies significantly across frameworks, software libraries, and hardware backends. Furthermore, statistical validation of these PRNGs is often neglected, which could open the door for adversarial manipulation of randomness.
The potential impact of compromised randomness can be profound, enabling adversaries to exploit these vulnerabilities covertly and go undetected. 
Prior work has demonstrated that attacks on randomness in machine learning systems can lead to degraded model performance on target classes~\cite{DropOutAttacks} or false robustness certifications~\cite{dahiyarandomness}.
Randomness quality also plays an important role in other areas such as cryptography, where researchers and practitioners have developed rigorous standards and constructions to ensure security. However, cryptographic requirements for randomness often differ from machine learning requirements which rely on specific distributions. Furthermore, cryptographic constructions impose significant overhead on ML systems which is practically untenable. As a result, machine learning systems require new threat models and standards as it relates to randomness.

%Previous research has demonstrated that attacks on randomness have been effective in real-world systems, including those targeting cryptographic applications and secure random number generation. In the context of ML, an adversary with control over the randomness sources could degrade model performance, induce unfair behavior, or even compromise privacy.

%\zahra{talk about how in this work we close this gap by conducting a comprehensive study of algos and impl of randomness in popular ml frameworks, and explore different attacker capabilities etc}

To address existing gaps, we perform a comprehensive study on sources and requirements for randomness in ML systems, explore attacker motivations and capabilities, and examine algorithms and practices in popular machine learning frameworks. Our study reveals that 
significant variations exist in design choices and implementations of randomness across different frameworks or even hardware backends within the same framework, with no consensus on policies to set seeds or third party libraries to use. Such variations often manifest in poor or insecure design choices such as setting a seed based on a fixed string or system time, or not using a cryptographically-secure generator with differential privacy, resulting in vulnerabilities that can be exploited by an attacker. 
Based on our ecosystem study, we propose a new taxonomy of attacks on ML systems targeting PRNGs and extract
%\todo{summary of survey results, eg impl of rand across os and hw backends are inconsistent even within the same framework with security implications, no standard way to set seeds, frameworks rely on different third party libraries. Notably, we identify a range of issues in popular frameworks ranging from obsolete and insecure practices such as using system time as randomness seed to implementing differential privacy with insecure algorithms. we also categorize attacks on randomness and mitigation techniques} 
%The output from this step is 
a set of policies dictating the requirements for randomness in ML systems. 
To address existing security gaps, we develop \sys, a system for securing randomness sources in machine learning frameworks. \sys consists of a static and a dynamic component. \sys's static component analyzes a library's source code and identifies instances of random functions and modules that use them.
At runtime, \sys enforces secure execution of random functions by replacing insecure function calls with \sys's implementations that meet policies derived from our study. \sys integrates two approaches to identify and mark insecure functions. In the first approach, \sys incorporates the policy verification as part of static analysis of the library. Alternatively, the second approach verifies policy requirements at runtime by running randomness quality tests in parallel.
The second approach provides more flexibility and can be extended to scenarios where access to source code is limited, but imposes a higher runtime penalty.

We evaluate \sys over PyTorch and demonstrate its capabilities to enforce secure randomness policies. Specifically, we evaluate \sys over (i) primitive RNG functions and (ii) an end-to-end machine learning pipeline in privacy-sensitive and privacy-insensitive settings. Our evaluations show that \sys static policy enforcement adds a moderate runtime overhead, and \sys dynamic policy enforcement with randomness tests adds a significant runtime overhead, which can be reduced through asynchronous and selective auditing optimizations. 

In this work, we make the following contributions: 
\begin{itemize}
\item We present an ecosystem study of ML frameworks and existing practices for generating and using random values as well as integrated quality tests. Our study exposes a range of issues including insecure policies for seed assignment and PRNG implementations. We use these results to derive policies for randomness in machine learning systems.
\item We present a novel attack taxonomy on ML randomness and explore attacker capabilities under different threat model assumptions. We use our model to classify existing work and categorize attacker objectives.
\item We design and implement \sys, a system for enforcing randomness policies. \sys uses static analysis to identify instances of random functions and modules that use them, and enforces secure execution at runtime. We evaluate \sys using an instantiation of the PyTorch library and demonstrate its effectiveness in practical settings.
%\item We evaluate \sys using an instantiation of the PyTorch library. Our results indicate that static enforcement of policies adds upto 47\% runtime overhead while dynamic enforcement incurs a significant overhead upto 780\% initially but reduces to 96\% after optimization across various training settings. 
\end{itemize}

\section{Background}
\label{sec:bg}
\subsection{Randomness in ML}
Randomness plays a crucial role in different stages of machine learning systems. 
Data processing involves splitting the dataset randomly to ensure that the train and test sets have the same distribution of classes and features. Randomness is also used in data augmentation where random (geometric, color, or kernel) transformations are applied to training data to improve model generalizability~\cite{imageaug}. 
%Weight initialization: 
Weight or parameters in a neural network are generally drawn randomly from the Gaussian or uniform distributions~\cite{GoodfellowDeepLearningBook}.

Optimization algorithms such as stochastic gradient descent (SGD) samples randomly selected subsets of the dataset in each round to compute gradients and update model weights.
Regularization techniques such as dropout~\cite{DropOutOG} and randomized smoothing~\cite{RSmoothingOG} are used during training to prevent overfitting by introducing random noise or randomly deactivating parts of the network. These strategies force the model to learn more robust features that generalize better to unseen data.
Randomness is also involved in other aspects of ML, such as in hyperparameter tuning strategies (random search) \cite{randomsearch}. 
Trustworthy ML axes such as privacy and robustness also use randomness in attacks and defenses. In particular, differential privacy (DP) \cite{dp} relies on the careful addition of random noise to the output of a computation to mask the inputs to that computation; the most common variant is $(\epsilon, \delta)$ DP---formally, an algorithm $\mathcal{M}$ satisfies $(\epsilon, \delta)$ DP if for all neighboring datasets (differing only by one element) $\mathcal{D}, \mathcal{D'}$ and all measurable $\mathcal{S}$, $P(\mathcal{M}(\mathcal{D} \in \mathcal{S}) \leq e^\epsilon P(\mathcal{M}(\mathcal{D'} \in \mathcal{S}) + \delta$. One of the ways to achieve this notion is through the addition of Gaussian or Laplacian noise \cite{dp}. Abadi et al. \cite{DPSGD} first proposed using DP for ML models by using Poisson sampling of batches to calculate gradients and adding Gaussian noise to them (DP-SGD), to hide private training data. 
% Federated learning devia
%%%% PRNG section -- 
\subsection{Pseudorandom Number Generators}
Properties of random sequences and methods to generate them have been the subject of extensive research. Ideally, true random numbers would be derived from random physical processes such as thermal noise, clock drift, and jitter. However, extracting random numbers solely from such sources is unrealistic for most applications since they don't provide sufficient throughput and processing them is time-consuming.
%The production of random numbers from real-world phenomenon is, however, time-consuming, making its use as the sole source of randomness unrealistic in most applications. 
%Therefore, most computing applications rely on pseudorandom number generators (PRNGs), deterministic algorithms that can generate seemingly random sequences of numbers quickly. 
PRNGs provide an alternative to generate random numbers efficiently.
Generally, a PRNG passes an initial seed value to a (deterministic) algorithm which then generates a sequence of random numbers.
%Generally, a PRNG uses an initial seed value to generate its first random number, and uses the previous number in the generated sequence or some kind of counter to generate the next random number. Most PRNGs have an internal state that is dependent on the initial seed, a function that updates the internal state and a function that maps the current state of the PRNG to a random number. Identical seeds will produce identical sequences of numbers. The output of PRNGs repeat after a certain sequence length (called period) has been generated; to avoid repetition, it is best practice to re-seed PRNGs. 
The most commonly used PRNG algorithms include the Mersenne Twister~\cite{MT19937}, Philox~\cite{PhiloxThreeFry}, and ThreeFry~\cite{PhiloxThreeFry}.

\mlparagraph{Cryptographic Security.}
In cryptography, random numbers are defined with precise requirements to ensure the security of various operations such as key generation, encryption, and authentication. We provide a brief description here and refer the readers to Katz and Lindell's work~\cite{yehuda} and the NIST 800-22 standard~\cite{NIST800} for more detail.
At a high level, a cryptographically secure PRNG (CSPRNG) is defined such that its output (with uniform distribution) is \emph{computationally indistinguishable} from a truly random sequence, i.e., an adversary will not be able to distinguish the output of CSPRNG from true randomness using any (polynomial time) algorithm. In practice, CSPRNGs are implemented by collecting a pool of entropy from sources of true randomness and using bits from the entropy pool as a seed in a cipher or hash function to generate random bits.

Given the initial seed, a sequence of random numbers generated by a PRNG can be computed entirely. As a result, in security-sensitive applications, the value of the initial seed should be secret and picked from a large range (to make brute-force attacks unfeasible) and PRNGs should be frequently reseeded. In Unix-like operating systems, true sources of randomness (thermal noise, fan noise, keyboard and mouse timings, etc.) are collected into an entropy pool and used to seed a CSPRNG, the output of which is available through \texttt{/dev/urandom}~\cite{devurandlinuxman}. Similarly, Windows-based systems maintain an entropy pool sourced from interrupt timings and use it to seed a CSPRNG~\cite{win10crypto}, the output of which is available via the \texttt{BCryptGenRandom} API. 
% Prior to Windows Vista, the now-deprecated \texttt{CryptGenRandom} API provided cryptographically secure random numbers \cite{cryptgen} and its implementation was largely unknown; the Windows 2000 implementation was reverse-engineered in 2012 by Dorrendorf et al. \cite{winvistarngbroken}. 

\subsection{Randomness from a Distribution}
\label{sec:randomnessfromdist}
The raw output of most PRNGs is a fixed-width integer (usually 64 or 32 bits) and interpreted to be a sample from a discrete uniform distribution, e.g., $\mathcal{U}(0,2^{64})$. 
%or $\mathcal{U}(0,2^{32})$. 
However, many ML algorithms need specific distributions (such as normal or uniform).
To this end, we can transform the raw output from a PRNG into the desired distribution as specified below.
\mlparagraph{Uniform Distribution.} We can transform a uniformly distributed integer sample $u \sim \mathcal{U}(0,2^{64})$ to a real-valued sample $z \sim \mathcal{U}(a,b)$ where $a,b \in \mathbb{R}$ by applying a linear mapping $z = {u}/{2^{64}}\cdot(b-a)+a$. In contrast, for an integer sample $z \sim \mathcal{U}(a,b)$ where $a,b \in \mathbb{Z}$ we use modular arithmetic to scale the output $z = u \mod{(b - a)} + a$.

\mlparagraph{Normal Distribution.} The Box-Muller transform \cite{boxmullerref} generates two normally distributed samples $z_0, z_1 \sim \mathcal{N}(0,1)$ from uniformly distributed samples $u_0, u_1 \sim \mathcal{U}(0,1)$ by computing $z_0 = \sqrt{-2\mathrm{ln}(u_0)} \cdot \mathrm{cos}(2\pi u_1)$ and $z_1 = \sqrt{-2\mathrm{ln}(u_0)} \cdot \mathrm{sin}(2\pi u_1)$.
We can transform any sample $z$ from a standard normal distribution  $\mathcal{N}(0,1)$ to a sample $x$ from a normal distribution $\mathcal{N}(\mu, \sigma^2)$ via scaling $ x \ \ = \ \  \sigma z \ \  + \ \  \mu$.

\mlparagraph{Laplace distribution.} We can generate a sample from the Laplace distribution $z \sim \mathcal{L}(\mu, b)$, using a uniformly distributed sample $u \sim \mathcal{U}(-1,1)$ via $ z \ \ = \mu - b\ \text{sgn}(u) \ln{(1+|u|)}$, where $\text{sgn(.)}$ is the signum function. 

%%%%%%%%%% Random Variable version ?
\subsection{Distribution Statistical Tests}\label{subsec:stataudit}
Validating the correctness or security of the underlying uniform PRNG is not sufficient to guarantee the correctness of distributional random numbers.
%The correctness of distributional transformations is not guaranteed by PRNG validation alone; even with a statistically sound PRNG, i
Incorrectly implemented transformation logic can result in deviations from the desired target distribution and requires dedicated test suites. Goodness-of-fit tests such as the Kolmogorov-Smirnov (KS)~\cite{KStest} and Pearson's $\chi^2$~\cite{PearsonChi} detect such deviations by estimating the empirical distribution of the sampled data and comparing it to the the target distribution. In this work, we provide a brief description of some tests here.

\mlparagraph{Z test.} is a statistical hypothesis test \cite{ztest} used to determine whether the mean of a sample is different from a known population mean. The test statistic is $Z = (\bar{x} - \mu)/(\sigma/\sqrt{n})$ where $\bar{x}$ is the sample mean, $\mu_0, \sigma$, the population mean and standard deviation, and n is the sample size. If the value of z is very large, it suggests that the sample mean is different from the given population mean. 

\mlparagraph{Kolmogorov-Smirnov test.} (KS test) is a distribution-independent test that measures the maximum distance (D-statistic) between the empirical cumulative distribution function (CDF) of the observed samples and the target distribution. The D-statistic is computed as $D = \sup_x(F_n(x) - F(x))$ where n is the sample size, $F_n$, the empirical CDF, $F$, the target CDF and $\sup_x$ is the supremum over the set of observations. The p-value is computed by $p = P_k(x > D_n)$ where $P_k(x)$ is the probability of observing $x$ from the Kolmogorov distribution. A small p value indicates a deviation from the target distribution.

\mlparagraph{Pearson's $\chi^2$ test.} (or $\chi^2$ test) is used to assess the goodness-of-fit between an observed frequency distribution and an expected discrete distribution. It uses the number of bins as its only parameter. The test statistic $\chi^2$ is computed as $\chi^2 = \sum_{i=1}^k\frac{(O_i - E_i)^2}{E_i}$ where $O_i$ and $E_i$ are the observed and expected frequencies of bin $i$, and $k$ is the number of bins. The p value is calculated similarly as above using the $\chi^2$ distribution. 

% \todo{maybe a brief description of dp auditing tools?}
%\subsection{Auditing Tools}
%DP auditing tools are used to verify the privacy guarantees and detect violations in differentially private mechanisms. They generally work by probing the DP mechanism with constructed inputs and analyzing the outputs for guarantee violations. First, they generate neighboring datasets (i.e. datasets that differ only in one entry), run the DP mechanism with those inputs, and test if the outputs of the DP mechanism differ more than they're allowed to, i.e. a privacy violation.  

\section{Threat model}
\label{sec:threatmodel}
In this section we provide a systematization of attacks on randomness in ML systems and present threat models which should be considered when developing mitigation techniques. At a high level, we consider two ways
in which randomness in ML systems can be compromised: 
\begin{enumerate}
    \item \textbf{Existing vulnerabilities:} Poor design choices or use of insecure algorithms that can lead to attacker exploitation.
    \item \textbf{Supply chain attacks:} An attacker deliberately compromising a library within the system to manipulate randomness sources.
\end{enumerate}

Exploiting existing vulnerabilities allows an attacker to compromise ML systems with relatively low effort. Examples include poor PRNG seeding policies or use of insecure (not cryptographically secure) PRNGs for DP. In the next section, we present a comprehensive study of open-source ML frameworks and identify vulnerabilities and corresponding attacks which malicious parties can exploit. We note that we followed responsible disclosure procedures to inform developers of discovered vulnerabilities in advance.

Supply chain attacks have been studied in broader software systems \cite{ladisataxonomy, softwaresupply} but have been underexplored in ML frameworks. Any piece of software relies on a supply chain of software components and an actor controlling any step in this chain may be able to sabotage downstream software.

By exploiting existing  the attacker some level of influence over the random operations used during training -- such as data shuffling, dropout, weight initialization, or augmentation. This influence can be leveraged to degrade model performance, or to target properties like privacy and fairness.

\textit{Why would an attacker manipulate randomness?} The primary goal for an attacker in this scenario is to degrade model performance and compromise guarantees like privacy and fairness. Such an adversary aims to go undetected, and therefore randomness provides a covert vector of attack that is often overlooked in ML. Other vectors of attack such as model serialization vulnerabiltiies~\cite{modelserialattack} have mitigation techniques~\cite{pickleball, modelscan}. However, randomness manipulation remains an unexplored attack vector with no existing solutions.

% In addition to analyzing a powerful attacker targeting ML frameworks, we also consider other attacker capabilities with less power such as the ability to only manipulate a third party library implementing the RNG or knowledge of the seed source as discussed in Section IV.C. We show that such an adversary can still compromise the system to achieve different objectives such as undermining privacy or robustness guarantees. 

\section{Randomness in ML Systems}
\label{sec:randml}
% \itodo{decide between star and bigstar}
%%%%%% Concretely talk about randomness in PyTorch framework, this is the ML pipeline Excel sheet. Classify based on pipeline as well.
%%%% Summary of statistics may be useful here\todo{can we reduce the width of the first column by writing the labels in two lines? data processing etc. requirements is the distribution except for first row, can we write it in terms of random permutation etc? merge attacks and effects columns into one as attacker capabilities and put citations there. what is attack vector? make sure columns are easy to understand from the description in the caption}

\begin{table*}[ht]
\caption{Sources of randomness in PyTorch \& ML. For each ML component, we list general techniques and the role and distributional requirements. We also present the potential objectives of an adversary seeking to exploit the technique. $n_i$ and $n_o$ denote the number of input and output channels of the layer. We indicate whether a technique used in a defensive context with $^\bigstar$. }
\label{table:MLpipelineRand}
\centering
\resizebox{\textwidth}{!} {
\begin{tabular}{p{2.5cm}p{3cm}p{3cm}p{3cm}p{5cm}}
\toprule
\textbf{Component} & \textbf{Technique} & \textbf{Randomness Use} & \textbf{Distribution}  & \textbf{Attacker objectives}\\
\midrule
\multirow{2}{*}{Data Processing} & Data Splitting & Train/test/val split & Uniform random permutation & Under/overfitting, unbalanced classes, lower accuracy \\
\cmidrule{2-5}
& Data Sampling for SGD & Batch selection & Uniform i.i.d. & Bad batches fed to SGD, unfair learning \cite{sgddata}\\
\midrule
\multirow{4}{*}{Data Augmentation} & Apply Transformations & Randomly transform & Uniform  &  \\
\cmidrule{2-4}
& Geometric (Crop, Flip) & Params for transforms (e.g., crop) & Uniform &  \\
\cmidrule{2-4}
& Colorspace  (ColorJitter) & Params for color transformations & Uniform  &Mislabel attacker-chosen inputs but preserve model accuracy\cite{rance2022augmentationbackdoors}   \\
\cmidrule{2-4}
& Kernel  (Gaussian Blur) & Random params (e.g., $\sigma$ for Gaus. blur) & Uniform & \\
\midrule
%%%%% Put glorot, nuniform, kaiming
\multirow{4}{4em}{Training} & \multirow{4}{*}{Weight Init.} & Uniform Xavier init. & $\mathcal{U}(\pm \sqrt{\frac{6}{n_i + n_o}})$ &   \\
\cmidrule{3-4}
 & & Normal Xavier init. & $\mathcal{N}(0, \sqrt{\frac{2}{n_i + n_o}})$  & Identical weights $\rightarrow$ same gradients, no learning; exploding gradient and vanishing gradient problem \cite{GoodfellowDeepLearningBook}; degrade model accuracy\cite{advweight}\\
\cmidrule{3-4}
 & & Kaiming init. & $ \mathcal{N}(0, \sqrt{\frac{2}{n_i}}) $ &  \\
\cmidrule{3-5}

& Fractional Maxpooling & Random regions for maxpooling & Uniform & Deliberate selection of pooling regions to cause poor generalization \\
\midrule
\multirow{2}{*}{Regularization} & Randomized Smoothing\cite{RSmoothingOG} $^\bigstar$ & Random noise $\epsilon$ added to input & $ \mathcal{N}(0, 1)$ & Inflated robustness metrics, increase certification time \cite{dahiyarandomness}\\
\cmidrule{2-5}
& Dropout \cite{DropOutOG}& Randomly drop neuron outputs & $\mathcal{U}$(0, 0.5) &  Exclusion of certain classes, unfair learning\cite{DropOutAttacks}\\
\midrule
\multirow{4}{*}{Differential Privacy $^\bigstar$} & Batching & Subsampling of batches & CSPRNG,  Poisson batch selection & \makecell[b]{--}  \\
\cmidrule{2-5}
 & Noise for SGD & Add noise to gradients & CSPRNG, Gaussian or Laplace & \\
 \cmidrule{2-4}
 & Central DP noise & Add noise to aggregated client updates & CSPRNG, Gaussian or Laplace&\makecell[b]{Recover gradients to reconstruct \\ private training data  \cite{deepleakage} }  \\
  \cmidrule{2-4}
  & Local DP noise & Add noise to client updates & CSPRNG, Gaussian or Laplace&  \\
\bottomrule
\end{tabular}
}
\end{table*}

In this section we perform a systematic study of randomness in ML systems as well as implementations in popular open-source frameworks. We start by exploring requirements that should be satisfied for each technique, including specifying distributional or security properties. Additionally, we will detail attacker goals targeting each technique's source of randomness. We then examine implementations in ML frameworks and analyze and compare supported algorithms, design choices, and identified vulnerabilities.

\subsection{Randomness Requirements and Attacker Objectives}\label{sec:randreq}

As described in Section~\ref{sec:bg}, randomness plays an important role in various components of ML systems summarized in Table~\ref{table:MLpipelineRand}. During data processing, randomness is commonly used to split datasets into training and validation subsets. These splits should be representative of the entire dataset and preserve class balance, generally following a uniform random permutation.  An attacker who targets randomness in data splitting may exclude specific samples or classes, leading to under/overfitting of the model and degraded metrics such as fairness.  
The data processing stage also involves preparing mini-batches for the Stochastic Gradient Descent (SGD) algorithm, requiring uniform sampling to compute an unbiased estimate of the gradient. Attackers targeting randomness for preparing batches may change the order of datapoints within batches or the order of batches themselves. Reordering attacks can result in reduced model accuracy, slowing down training or resetting progress, injection of backdoors~\cite{sgddata}, and pushing the model towards unfair behavior by affecting group-level accuracy and fairness~\cite{mlfairness}.

% The data processing stage typically involves randomly splitting the dataset into subsets (for example training and validation). The splits should be representative of the entire dataset and preserve balance across classes, generally following a uniform random permutation.  An attacker targeting randomness for data splitting may aim to exclude specific samples or classes, leading to under/overfitting of the model and imbalanced classes affecting metrics such as fairness. 
% Data processing also involves preparing batches to be consumed by the Stochastic Gradient Descent (SGD) algorithm, requiring a uniform distribution to compute an unbiased estimate of the gradient. Attackers targeting randomness for preparing batches may aim to change the order of datapoints inside batches or change the order of batches themselves. Reordering attacks alone can result in reduced model accuracy, slowing down training or resetting progress, injection of backdoors~\cite{sgddata}, and pushing the model towards unfair behavior by affecting group-level accuracy and fairness~\cite{mlfairness}.
Randomness is also used in data augmentation operations, where random transforms increase the training data diversity and reduce overfitting \cite{imageaug}. Common augmentations include geometric transforms (e.g. scaling, rotation), colorspace transforms (e.g. changing brightness, contrast, saturation) and kernel transforms (e.g. Gaussian blur). These transformations typically sample parameters from a uniform distribution. If an attacker compromises the randomness source, they can insert stealthy backdoors that are hard to detect and preserve model accuracy. In such cases, the model behaves normally on most inputs but produces attacker-chosen outputs for specific inputs at inference, leading to decreased class-level accuracy \cite{rance2022augmentationbackdoors}. 
% Data augmentation steps apply different transforms on data points in order to increase the training data diversity and reduce overfitting \cite{imageaug}. Geometric transforms change the spatial orientation of an image, such as scaling and rotation. Colorspace transformations adjust the color representation by changing brightness, contrast, saturation etc. Kernel transforms like Gaussian blur apply convolutional operations to images. The randomness in the aforementioned augmentation techniques is sampled from a uniform distribution. An attacker targeting the randomness source in data augmentation can insert backdoors that are hard to detect and preserve model accuracy. In such cases, the model behaves normally on most inputs but produces attacker-chosen outputs for specific inputs at inference, leading to decreased class-level accuracy \cite{rance2022augmentationbackdoors} 

Before training begins, the weights of the model are initialized to random values in order to \emph{break symmetry}~\cite{GoodfellowDeepLearningBook} and learn distinct features. Common weight initialization techniques include uniform and normal variants of Xavier (Glorot) initialization~\cite{XavierInit} and Kaiming initialization~\cite{KaimingInit}.
Attacks on weight initialization~\cite{advweight} require limited knowledge of the model architecture but do not require any knowledge of the training data. Such attacks can subtly degrade the model's performance by altering or shifting weight matrices resulting in reduced accuracy or increased training time~\cite{advweight}. Apart from weight initialization, randomness also appears in fractional maxpooling \cite{graham2015fractionalmaxpooling} layers that perform fractional downsampling by randomly choosing pooling regions. Fractional maxpooling has reduces overfitting \cite{graham2015fractionalmaxpooling} in convolutional neural networks.  While no attacks have been noted, an attacker could bias the pooling region selection or manipulate the downsampling ratio to nullify its benefits. 

% Attacks targeting weight initialization do not need access to the dataset and requires limited knowledge of the model architecture. However, such attacks \cite{advweight} can subtly degrade the model's performance by altering or shifting weight matrices or changing the proportion of small and large weights resulting in reduced accuracy or increased training time. Similarly, the attacker can periodically shift the columns of the weight matrices after initialization to deactivate certain neurons, leading to degraded model accuracy without affecting the behavior of the loss.  Apart from weight initialization, fractional maxpooling \cite{graham2015fractionalmaxpooling} layers are a variation of the maxpooling layers that allow for the fractional downsampling of input to the layer by randomly choosing pooling regions, and has been found to reduce overfitting \cite{graham2015fractionalmaxpooling} in convolutional neural networks.  While no attacks have been noted, an attacker may bias or override the region selection and manipulate the downsampling ratio to nullify its benefits. 

%When DP is used in the federated learning setting,  it aims to limit the information leaked from a client's gradient update by the addition of noise; an attacker might aim at reversing each client's noised gradients to learn about their training data through gradient leakage attacks \cite{deepleakage}. 

Finally, randomness is central to robustness and privacy mechanisms. Randomized smoothing~\cite{RSmoothingOG} is a certification technique that provides provable guarantees on model robustness by adding Gaussian noise to input data and aggregating predictions across noisy samples. An attacker can alter the certification noise to produce false robustness guarantees \cite{dahiyarandomness} and make a model appear certifiably robust when it is not. Dropout~\cite{DropOutOG} randomly masks activations during training to prevent overfitting, and an attacker who controls the dropout masks can bias learning outcomes and make the model unfair~\cite{DropOutAttacks}. In privacy settings, DP in ML requires the DP-SGD noise to be generated using a CSPRNG~\cite{HolohanIBM}.

% Randomized Smoothing (RS) \cite{RSmoothingOG} is a certification technique in machine learning that helps certify model robustness by adding Gaussian noise to input data and evaluating model predictions over noised versions of a given input; RS provides provable guarantees on model robustness within a certain radius of the input (certification radius). An attacker can alter the certification noise to produce false robustness guarantees \cite{dahiyarandomness} and make a model appear certifiably robust when it is not. Dropout \cite{DropOutOG} is a regularization technique that  randomly zeroes out some activations during training to prevent overfitting. Attacks targeting dropout \cite{DropOutAttacks} aim at controlling the dropout masks to exclude certain datapoints or classes, making the model unfair. 
% DP in ML settings require the DP-SGD noise to be generated using a CSPRNG \cite{GarfinkelLeclerc}. 
\subsection{Ecosystem Study}
\label{sec:implframe}
We conduct an ecosystem study to understand the design of PRNG across different ML systems and evaluate and compare security, seeding practices, and integrated tests. The summary of our findings is included in Table~\ref{table:impl} which covers popular ML frameworks, DP libraries, and open-source federated learning platforms incorporating DP. For each framework, we present the supported PRNG algorithms, seeding practices, and whether the components are natively implemented or depend on third-party libraries. By comparing our study with the randomness requirements derived from Section~\ref{sec:randreq}, we can identify implementation issues, security vulnerabilities, and potential supply chain attack vectors as we will discuss next. 

%We look at some common ML Frameworks, other popular Python libraries that have PRNG implementations and differential privacy for machine learning (DP-ML) libraries in Table \ref{table:impl}. We also classify the PRNGs based on whether they are natively implemented or depend on third-party libraries, accounting for potential supply-chain threat vectors. Our survey considers whether the PRNGs are cryptographically secure -- especially relevant in DP settings -- in addition to initialization of seeds, and user ability to set seeds, which has important implications for reproducibility and seed manipulation. 

\subsubsection{ML Frameworks} We start by describing popular ML frameworks and emphasize any issues discovered.

\mlparagraph{PyTorch}~\cite{PyTorchCitation} uses two different natively-implemented PRNGs depending on the hardware backend; it uses MT19937~\cite{MT19937} on the CPU and Philox~\cite{PhiloxThreeFry} on the GPU. Both implementations use OS-provided randomness on all platforms (utilizing \texttt{/dev/urandom}). However, a notable exception is for 32-bit Windows systems, which instead uses system time for seeding. System time is not a secure source of entropy, as it can be predictable or recoverable by an attacker. 

\mlparagraph{JAX}~\cite{JAX} emphasizes reproducible code and as a consequence requires that all seeds be provided by the user. JAX supports both natively-implemented or XLA-accelerated versions of ThreeFry and Philox algorithms~\cite{PhiloxThreeFry}. 

\mlparagraph{TensorFlow}~\cite{Abadi_TensorFlow_Large-scale_machine_2015} uses Philox by default on all hardware backends and optionally supports ThreeFry through XLA (Accelerated Linear Algebra), a compiler that optimizes ML code~\cite{XLA}. TF seeds its PRNGs from system entropy via TSL (Tensor Standard Libraries), which provides low-level implementations for TensorFlow functions \cite{TSL}. 

\mlparagraph{Keras}~\cite{Keras} is primarily an API layer that can be used with different frameworks. However, Keras implements specific seeding practices independent of the backend framework. By default, the seed value is set to the string \texttt{global\_seed} which can be fed into PRNGs from TensorFlow or JAX.

\mlparagraph{TF$.$Keras}~\cite{tfkeras} is the TensorFlow-specific implementation of Keras which draws its seed uniformly from $[1, 10^9] $ using Python's \texttt{random} module (based on Python-implemented MT19937). Python's \texttt{random} either uses OS-provided randomness (\texttt{/dev/urandom}) or falls back on system time and process ID for its default seed if the former is unavailable.

\subsubsection{ML-related Libraries} We extend our study to include popular libraries that are widely used in developing ML applications directly or as dependencies.

\mlparagraph{NumPy}~\cite{numpy} provides several PRNG implementations based on PCG64~\cite{PCG64}, MT19937~\cite{MT19937}, Philox~\cite{PhiloxThreeFry} and SFC64~\cite{sfc} and draws its default seeds from Python's {secrets} library. Even though Python's {secrets} generates seeds using a CSPRNG \cite{pythonSecrets}, the PRNG algorithms themselves are not cryptographically secure. . 

\mlparagraph{scikit-learn}~\cite{scikit-learn} is a ML library built on top of SciPy~\cite{scipy} and relies entirely on NumPy's PRNGs for all of its stochastic operations and does not support cryptographically secure randomness.

\subsubsection{DP Libraries} We study several popular DP libraries and discuss security guarantees provided by implementations.

\mlparagraph{Opacus}~\cite{opacus} is built on PyTorch and allows for the use of a CSPRNG but defaults to insecure PRNGs due to the overhead imposed by using a cryptographically-secure PRNG. Accordingly, Opacus recommends using the insecure mode during development and then switching to the secure mode for the final training and model deployment. The CSPRNG implemented in Opacus, namely {torchcsrpng}, is the only CSPRNG implementation that is supported by PyTorch. {torchcsprng} implements AES-CTR CSPRNG~\cite{PhiloxThreeFry} using \texttt{/dev/urandom} as its seed source and supporting both CPU and GPU hardware backends. However, the torchcsprng library has not been updated since 2021 and is not compatible out of the box with newer versions of Opacus and PyTorch. 

\mlparagraph{TensorFlow Privacy}~\cite{tfprivacy} (TFPrivacy) inherits TensorFlow's PRNGs but does not offer CSPRNGs for their DP-SGD optimizer. Instead, TFPrivacy uses Gaussian noise for central DP aggregation methods with TensorFlow's PRNGs using a system time seed as default.

\mlparagraph{Diffprivlib}~\cite{diffprivlib} uses Python's {secrets} library directly for its randomness rather than using it as a seed to a CSPRNG. Diffprivlib also offers NumPy's PRNGs as an (insecure) alternative.

\mlparagraph{JAX Privacy}~\cite{jaxprivacy} relies on JAX's PRNGs and therefore does not offer any CSPRNGs. Furthermore, JAX Privacy requires users to set seeds explicitly.

\subsubsection{Federated Learning (FL) Platforms} We also include two open-source federated learning libraries in our study that include DP components:

\mlparagraph{FedML-AI}~\cite{fedmlai} is a federated learning framework that uses PyTorch's PRNGs for its randomness. For central DP noise, FedML-AI uses \texttt{torch.normal} with the default options, i.e. a MT19937 seeded by \texttt{dev/urandom}; we note that the user cannot change the PRNG without modifying source code. 

\mlparagraph{TensorFlow Federated}~\cite{tffederated} (TFFederated) provides federated learning with central DP aggregation as well as local DP. TFFederated relies on TFPrivacy's aggregation methods, which relies on system time. System-time seeding can be particularly problematic as it provides low entropy and predictable seeds and has been exploited in the past to break encryption protocols\cite{systemtimeatk}. 

\begin{table*}[ht]
\caption{Implementation of PRNGs in ML frameworks \& third-party libraries. $\star$ indicates that the PRNG algorithm is default. We list supported algorithms, dependencies (e.g., XLA, TSL), default seeding practices (e.g., system time, OS entropy), whether users can change seeds, and if a CSPRNG is available for each framework. \yes{2pt} indicates availability and \no{2pt} indicates unavailability. }

\label{table:impl}

\begin{minipage}{2\columnwidth}
\begin{center}
  
\begin{tabular}{llllcc}
\toprule
\thead{\textbf{Framework}} & \thead{\textbf{{PRNG}}} & \thead{\textbf{{Implementation}}} & \thead{\textbf{Default seed}} & \thead{\textbf{User-defined} \\ \textbf{seed}}&   \thead{\textbf{CSPRNG}} \\
\midrule
\multirow{3}{*}{{PyTorch (PT)}} & \multirow{3}{*}{{MT19937 (CPU)$^\star$}} &  & \makecell[l]{Intel SGX: sgx\_read \\$\rightarrow$ SGX tRTS} & \yes{2pt} & N/A  \\
           &       & Native  & Linux: dev/urandom & \yes{2pt}     & N/A        \\
            &       &  & Windows: System Time, Process ID & \yes{2pt}   &        \\
\cmidrule{2-6}
% & & & & & \\
& Philox (GPU)$^\star$   &Native & \makecell[l]{CUDA: std::random\_device \\  $\rightarrow$ dev/urandom} & \yes{2pt}  &  N/A                                                   \\
\midrule
\multirow{3}{*}{{TensorFlow (TF)}} & ThreeFry & \makecell[l]{ XLA: XLA $\rightarrow$ TSL \\ $\rightarrow$ C++ STD, dev/urandom }&  & \yes{2pt}   &                                   \\
& & & \makecell[l]{TSL random $\rightarrow$ C++ STD \\, dev/urandom}  & & N/A \\

 & Philox $^\star$  &  \makecell[l]{non-XLA: TSL $\rightarrow$ \\ C++ STD, dev/urandom} &  & \yes{2pt}                                & \\
 
\midrule
\multirow{2}{*}{{JAX}} & ThreeFry$^\star$ &\makecell[l]{XLA   $\rightarrow$ TSL}  & User & \yes{2pt}   &  N/A  \\

& Philox  & XLA   $\rightarrow$ TSL & User & \yes{2pt}   &  N/A                                                    \\
\midrule
TF.Keras \footnote{tf.keras is Keras used with a Tensorflow backend} &  TF PRNG $^\star$  &TF & \makecell[l]{Seed drawn $[1, 10^9]$ via\\ Python random $\rightarrow$ dev/urandom} & \yes{2pt} & N/A \\
\midrule
Keras\footnote{Keras is framework-agnostic and can be used with ML frameworks like PyTorch, TensorFlow, JAX etc. }  &  Framework   & Framework & \makecell[l]{Seed set to string \\ "global\_seed"}&\yes{2pt} & N/A \\
\midrule
\multirow{5}{*}{{NumPy}} & PCG-64$^\star$& &   & & \\
                            & PCG-64 DXSM  &   & \makecell[l]{Linux: Python secrets \\ $\rightarrow$ dev/urandom}  &     &       \\
                            &MT19937 &  Native& &\yes{2pt}  &N/A  \\
                           & Philox&   &  \makecell[l]{Windows: Python secrets \\ $\rightarrow$ BCryptGen} & & \\
                           & SFC64 &   &  &   &  \\
\midrule
 scikit-learn & NumPy$^\star$ &NumPy & NumPy & \yes{2pt} & N/A\\
\midrule
\multirow{2}{*}{{Opacus}} &  AES-CTR  & torchcsprng $\rightarrow$ PT &dev/urandom& \yes{2pt}&\yes{2pt} \\
     & PT PRNG $^\star$& PT & PT &   \yes{2pt} & \no{2pt}           \\
\midrule
\multirow{2}{*}{TFPrivacy}  & TF PRNG & TF &  DP-SGD:  Refer TF &\yes{2pt} & \no{2pt}\\
& & &Central DP: System Time& & \\
\midrule

\multirow{2}{*}{{Diffprivlib}} & Python secrets & \makecell[l]{Linux: Python secrets \\ $\rightarrow$ dev/urandom \\Windows: Python secrets \\ $\rightarrow$ BCryptGenRandom}  & N/A & \no{2pt} & \yes{2pt} \\
\cmidrule{3-6}
  & NumPy PRNG $^\star$& NumPy  & NumPy &   \yes{2pt} & \no{2pt} \\
\midrule

JAX Privacy & JAX PRNG$^\star$ & JAX & User & \yes{2pt} &\no{2pt}\\
\midrule

FedML-AI & PT PRNG$^\star$ & PT & PT & \no{2pt} & \no{2pt}\\
\midrule
TFFederated & TF PRNG$^\star$ & TF & TF Privacy & \yes{2pt}  & \no{2pt} \\
\bottomrule
\end{tabular}
\end{center}
\end{minipage} 
\end{table*}

\subsubsection{Discussion}
\label{sec:ecosystemdiscussion}
% \itodo{discuss observations here. mention varying practices, supported algorithms, and dependencies. discuss seeding practices based on system time or user input or fixed string and why they may not be a good idea. relate this to attacks discussed previously if possible (both dp and non-dp) they should be fixed or at least warnings be provided. talk about csprng in dp and fl libraries and mention which ones are actually secure and which are not}
Our ecosystem-wide study reveals significant variation in how ML Frameworks and libraries implement and seed
%, and verify 
PRNG implementations. Some frameworks such as PyTorch implement their own PRNG algorithm natively, while others like JAX and TensorFlow have to rely on third-party libraries like XLA which in turn depends on TSL. Such layered dependencies makes it difficult to completely verify randomness sources and exposes the framework to covert supply-chain attacks. Similarly, seeding practices are different across the board, with some frameworks falling back to insecure defaults like system time and fixed strings. As previously discussed, seeds based on system time can be predictable or reverse-engineered by an attacker. Obtaining the seed to a PRNG will enable an attacker to derive the entire randomness sequence trivially, and as a result, obtaining initial model weights, batch order, and dropout masks leading to attacks previously discussed~\cite{advweight, sgddata, DropOutAttacks}. 
%furthermore, with fixed-string seeds (as in Keras\cite{Keras}), the attacker can obtain the PRNG noise trivially. 
The JAX ecosystem \cite{jaxprivacy, JAX} exclusively relies on user-supplied seeds. While this practice can be useful for reproducibility, it completely shifts the security responsibility to the user and might endanger the security posture of the system if users have little security expertise.
Furthermore, most of the DP-ML frameworks we surveyed reused their base framework's PRNG for noise addition, which may itself have insecure practices (as discussed above). Federated Learning libraries like TFFederated and FedML-AI are particularly vulnerable; prior knowledge of the seed could be used to denoise gradient updates from clients, and make it possible to leak sensitive client-side training data \cite{deepleakage}. Even though CSPRNGs are a strict requirement for DP and DP-SGD \cite{GarfinkelLeclerc, HolohanIBM}, they are inconsistently adopted across frameworks. Among the frameworks studied, only Opacus and Diffprivlib make use of CSPRNG noise via torchcsprng and Python {secrets}. As noted before, torchcsprng is incompatible with current Opacus and PyTorch versions, limiting support to older hardware backends. Moreover, users may miss out on security-relevant updates in Opacus and PyTorch, particularly those affecting noise generation. Other libraries like TFPrivacy and JAX Privacy rely on insecure PRNGs from TensorFlow and JAX and do not warn the user of insecure noise generation. 

%\zahra{we don't need to include this in the paper}
% \todo{\mlparagraph{Response from the disclosure process: } We inform the developers of each framework for which our survey identifies randomness-related issues. We only received a response from JAX Privacy developers, who acknowledge the issue and indicate that they plan to update their documentation to warn users about predictable seeds. Other libraries (such as TensorFlow Privacy) are aware of the lack of cryptographically-secure PRNG and leave it as future work~\cite{mcmahan2019generalapproachaddingdifferential} As a result, there is no confirmed timeline for fixes to be available in open-source libraries and they are applied at developers’ discretion who may have limited bandwidth, possibly leading to potential exploits.}

\begin{table}[ht]
\centering
\caption{Attack taxonomy based on capabilities of the attacker in targeting seed source, library dependencies, or the ML framework via exploiting existing vulnerabilities or performing supply chain attacks. Attack objective can be to compromise (a) performance, (b) privacy, (c) fairness, (d) integrity (such as inserting backdoors), and (e) robustness. }\label{tab:attacktax}
\begin{tabular}{lccc}
\toprule
Attack Target & \begin{tabular}[c]{@{}c@{}}Existing\\ Vulnerabilities\end{tabular} & \multicolumn{1}{c}{\begin{tabular}[c]{@{}c@{}}Supply \\ Chain Attacks\end{tabular}} & \begin{tabular}[c]{@{}c@{}}Attack\\ Objective\end{tabular} \\
\midrule
Seed source & \begin{tabular}[c]{@{}c@{}}Constant\\ System time\\ Limited range\\ User defined\end{tabular} & \begin{tabular}[c]{@{}l@{}}Manipulate\\ seed value\end{tabular} & (a)-(c) \\
\midrule
Dependencies & \begin{tabular}[c]{@{}c@{}}Insecure seed\\ value and PRNG\end{tabular} & \begin{tabular}[c]{@{}l@{}}Manipulate\\ seed value \\ and PRNG\end{tabular} & (a)-(e) \\
\midrule
ML framework & \begin{tabular}[c]{@{}c@{}}Insecure seed\\ value and PRNG\end{tabular} & \begin{tabular}[c]{@{}l@{}}Manipulate\\ seed value,\\ PRNG, and\\ higher func\end{tabular} & (a)-(e)
\\\bottomrule
\end{tabular}
\end{table}

\subsection{Attack Taxonomy}
We present an attack taxonomy based on our ecosystem study (Section~\ref{sec:implframe}) and threat model (Section~\ref{sec:threatmodel}). We identify three attack targets: (1) seed source, (2) library dependencies, and (3) ML framework. For each target, we summarize existing vulnerabilities and possible supply chain attacks which can target privacy, fairness, integrity, robustness, and performance as described in Table~\ref{tab:attacktax}. 

%Under this general threat model, we identify three ways in which an attacker can exert control: (1) the ability to predict and manipulate seeds, (2) the injection of a malicious dependency in the ML supply chain, and (3) load a malicious ML framework. These capabilities reflect varying degrees of control, as the attacker may not have full access to randomness in the training process. The objective of the adversary as a whole is to degrade model performance and target guarantees like privacy and fairness. We classify existing attacks on ML randomness in the context of this threat model.

\mlparagraph{Seed Source.} An attacker targeting the PRNG seed may have the ability to predict or manipulate the seed values. As discussed in Section~\ref{sec:ecosystemdiscussion}, existing vulnerabilities in this category consist of constant seed values, using system time or a limited range for deriving the seed, or relying on user defined seeds without security enforcing mechanisms. Supply chain attacks on the seed source can target underlying dependencies (such as compromising the call to \code{/dev/urandom} at the system level or using a malicious \code{secrets} library). As discussed previously, the ability to predict or manipulate the PRNG seed can result in complete breakdown of privacy guarantees provided by DP or result in unfair models. 

%\todo{
\textit{Example: } TFFederated implements DP mechanisms (including central DP aggregation methods) via TFPrivacy~\cite{tfprivacy}. In this setting, the PRNG seed used to generate DP noise is derived from server time and is incremented deterministically each round. A malicious FL client can observe noisy global updates and approximate the server start time, which constrains the seed to a small search window. At $1 \ \mu s$ precision, a $10 \ s$ time window corresponds to only 23 bits of entropy, which can be brute-forced offline. The attacker can deterministically reconstruct the per-round noise values and obtain denoised gradient estimates, undermining DP guarantees and enabling downstream reconstruction attacks~\cite{deepleakage}. 
%}

\mlparagraph{Library Dependencies.} An attacker targeting library dependencies can similarly exploit their existing vulnerabilities or trick the user into downloading a compromised version of a library, repository, or software package used by an ML framework. Insecure seeds or PRNG implementations (i.e., not cryptographically secure for DP) in existing dependencies can be exploited by attackers which may be harder to audit or verify. Similarly, supply chain attacks on dependencies can compromise the security of PRNGs (and their seed). Compromising library dependencies do not need the attacker to alter any part of the development cycle or training steps and can give the attacker a wide range of capabilities to affect performance, privacy, fairness, integrity, and robustness.

%\todo{
\textit{Example: } Dahiya et al.~\cite{dahiyarandomness} show that an attacker can manipulate the randomness in a third-party library like NumPy~\cite{numpy} used by randomized smoothing~\cite{RSmoothingOG} via a malicious PRNG to sabotage robustness guarantees.
%}

\mlparagraph{ML Frameworks.} Similar to insecure dependencies, existing vulnerabilities in ML frameworks consist of insecure seeds or PRNG algorithms. On the other hand, supply chain attacks on ML frameworks provide the strongest attacks, giving the malicious actor the ability to not only manipulate seeds and PRNG algorithms, but also override higher functions and possibly parts of the ML pipeline. However, such attacks are less covert than previous two categories.

%\todo{
\textit{Example: } The dropout attack~\cite{DropOutAttacks} selectively drops neurons during training by manipulating the dropout layers within their framework-level implementation, thereby affecting class-level accuracy and fairness guarantees.

\subsection{Existing Defenses}
In this section, we explore existing defenses against attacks on randomness, including available modules or auditing tools. 

\begin{table*}[ht]
\caption{Distributions  available in ML Frameworks and implemented defenses. Baseline checks indicates the type of framework tests implemented as described in Section~\ref{subsubsec:unittests}. We also report whether frameworks implement any statistical tests, and list the tested distributions, if any. \no{} indicates unavailability of tests. }

\label{table:frameworktests}
\begin{minipage}{2\columnwidth}
\begin{center}
  
\begin{tabular}{lllll}
\toprule
\textbf{Framework} & \textbf{Baseline checks}  &\textbf{Statistical tests} & \textbf{Distributions}   &     \textbf{Tested Distributions}                  \\

 \midrule
PyTorch & Reproducibility, Correctness, Property &\no{2pt}\footnote{PyTorch has implemented $\chi^2$ test only for the torch.Distributions module, which is not used by any ML functionality} & Gaussian, Uniform   & N/A   \\
\midrule

TensorFlow & Reproducibility, Correctness, Property & Z-test & \makecell[l]{Bernoulli, Binomial, Beta, Categorical \\ Dirichlet, Exponential, Gamma \\Laplace, Gaussian, Multinomial, Uniform}  & Binomial, Gamma \footnote{only if used with Keras} \\
\midrule

JAX & Reproducibility, Correctness & \no{2pt} &  \makecell[l]{Binomial, Beta, Categorical \\ Exponential, Gamma \\Gaussian, Multinomial, Uniform\\ Poisson, Student's t, Weibull, Orthogonal \\ Lognormal }& N/A \\
\midrule

TF.Keras & Reproducibility, Correctness & \no{2pt}& TF& N/A\\

\midrule

Keras & Reproducibility, Correctness, Property & \no{2pt}& \makecell[l]{Binomial, Beta, Categorical \\ Exponential, Gamma \\Gaussian, Multinomial, Uniform} & N/A\\
\midrule
NumPy &Reproducibility, Correctness, Property &\no{2pt}& \makecell[l]{Beta, $\chi^2$, Cauchy, Laplace,\\ Exponential, Gamma,  \\Gaussian, Uniform }& N/A \\
\midrule

scikit-learn & N/A& \no{2pt} & N/A  & N/A \\
\midrule

Opacus & Reproducibility, Correctness & \no{2pt} & Gaussian & N/A\\
\midrule

TFPrivacy &Reproducibility, Correctness &\no{2pt} & Laplace, Gaussian & N/A \\
\midrule

Diffprivlib & Reproducibility, Correctness, Property & \no{2pt} & \makecell[l]{Bingham, Exponential \\ Gaussian, Geometric \\ Laplace, Uniform}  &N/A \\
\midrule

JAX Privacy & Correctness & KS test & Gaussian & Gaussian\\
\midrule

FedML-AI & N/A & \no{2pt} & Laplace, Gaussian & N/A\\
\midrule

TFFederated & Property & \no{2pt}& Gaussian & N/A\\
\bottomrule

\end{tabular}
\end{center}
\end{minipage} 

\end{table*}

\mlparagraph{PRNG Statistical Tests.} The quality of a PRNG can be measured and validated using statistical tests.
TestU01 suite by L'ECuyer and Simard~\cite{TestU1}, SP800-22 suite by the National Institute of Standards and Technology (NIST)~\cite{NIST800}, and SP800-90 Series of standards by NIST~\cite{SP80090} are among the most widely-known and practical test suits to assess the quality of a PRNG output sequence.
These test suites examine whether the PRNG output deviates from a uniformly random sequence (or a truly random sequence) using hypothesis testing. For instance, the MonoBit test in the SP800-22 suite measures the proportion of ones and zeros in a bitstream; a truly random sequence would have approximately the same number of zeros as ones and any sequence that does not obey this would fail the test. As noted by NIST, while the statistical tests should not be used as the standard to certify a PRNG, it can be used as a first step in deciding whether to use that PRNG~\cite{NIST800}. To close this gap, the SP800-90 series provides a comprehensive framework for validating CSPRNGs including specific requirements for entropy sources, seeding practices, and validation of entropy sources~\cite{SP80090}. 
To our knowledge, none of the ML frameworks we study run their PRNG implementations through a battery of statistical tests nor do they report these results.

\mlparagraph{ML Framework Tests.}
\label{subsubsec:unittests}ML frameworks such as TensorFlow, PyTorch, and others often include built-in functionality to handle randomness, and some frameworks have implemented various tests to validate randomness-related aspects. These tests primarily aim to ensure reproducibility, uniformity in data sampling, and the quality of random number generation. However, these built-in tests are lightweight unit tests rather than statistical evaluations; they generally do not evaluate whether randomness remains unbiased or secure during program execution or across different hardware backends. In Table~\ref{table:frameworktests}, we provide a summary of supported distributions and corresponding tests in each ML framework we study. Available tests include 
%In the frameworks we surveyed in table \ref{table:frameworktests}, these 
baseline checks testing for \emph{reproducibility} (ensuring the same seed consistently produces the same output), \emph{correctness} (verifying that a given seed yields the expected output according to the PRNG algorithm and it lies within expected bounds), and \emph{distributional properties}. 
As Table~\ref{table:frameworktests} shows, 
%\mlparagraph{Distributional tests} Table \ref{table:frameworktests} also shows the statistical tests (if any) implemented in ML frameworks. Our findings reveal that 
most ML frameworks only implement some baseline checks  for reproducibility or correctness but do not statistically verify their distributional transformations. Among all frameworks, JAX Privacy and PyTorch (albeit only for a particular submodule that is not used by any ML functionality) are the only tools that test randomness with distribution statistic tests. TensorFlow applies statistical tests inconsistently and only for some distributions when used via Keras.

\mlparagraph{DP Auditing Tools.}
\label{para:dpaudit}
DP auditing tools are used to verify the privacy guarantees and detect violations in differentially private mechanisms. 
DP auditing tools such as DeltaSiege~\cite{DeltaSiege} typically operate by first generating a neighboring dataset (i.e. datasets that differ only in one entry), running the DP mechanism, and testing if the outputs of the DP mechanism violate privacy bounds for a given $(\epsilon, \delta)$. A privacy violation occurs if for some pair of inputs, $a, a'$, and a given $\epsilon, \delta$-private algorithm $\mathcal{M}$, and the set of all possible outputs of $\mathcal{M}$,  $\mathcal{S}$, $P(\mathcal{M}(a) \in \mathcal{S}) > e^\epsilon P(\mathcal{M}(a') \in \mathcal{S}) + \delta$. DP auditing tools \cite{DeltaSiege, DPAuditorium} rely on repeated queries or estimations to empirically calculate these probabilities. Nevertheless, DP auditing tools are not able to find privacy violations due to insecure PRNG use as we demonstrate later in this paper.

\subsection{Mitigation Techniques}
Based on previous discussion on existing defenses in ML frameworks to detect attacks on PRNGs, we conclude that new mitigation techniques are needed to protect ML systems from malicious attackers. In this section, we outline our solution to address existing security gaps. Based on our threat model and attack taxonomy, we consider both existing vulnerabilities and supply chain attacks as vectors that can be exploited by adversaries. 
While software supply chain security solutions such as in-toto~\cite{intoto} and Sigstore~\cite{sigstore} can be directly integrated in ML systems to protect frameworks and dependencies from malicious modifications, such solutions do not apply to vulnerabilities that arise due to unintentional (or even malicious) insecure implementations within ML frameworks and libraries.
%\todo{
On the other hand, patching each framework or creating a secure fork may seem like an intuitive approach but it is not a viable solution in practice. Developers of open-source ML libraries may not have the bandwidth to apply fixes in a timely manner even when they are aware of existing issues. Similarly, creating a secure fork introduces additional overhead of maintaining the fork. As a result, upstream developer patches alone are insufficient and can leave downstream users vulnerable. 
%}

To address this gap, we propose \sys, a system designed to protect applications that use ML libraries against insecure PRNG practices and implementations. We will present the design of \sys next, followed by performance evaluations to demonstrate its practicality.

\section{\sys}
We introduce \sys, a system for securing randomness sources in ML systems as described next. 
%\sys consists of a static and a dynamic component as depicted in Figure~\ref{fig:sys} and described next. 
%that enforces secure and auditable randomness in ML pipelines. \sys is designed to mitigate both accidental vulnerabilities and targeted attacks on randomness mechanisms.
\begin{figure}[t]
    \centering
\includegraphics[width=\linewidth]{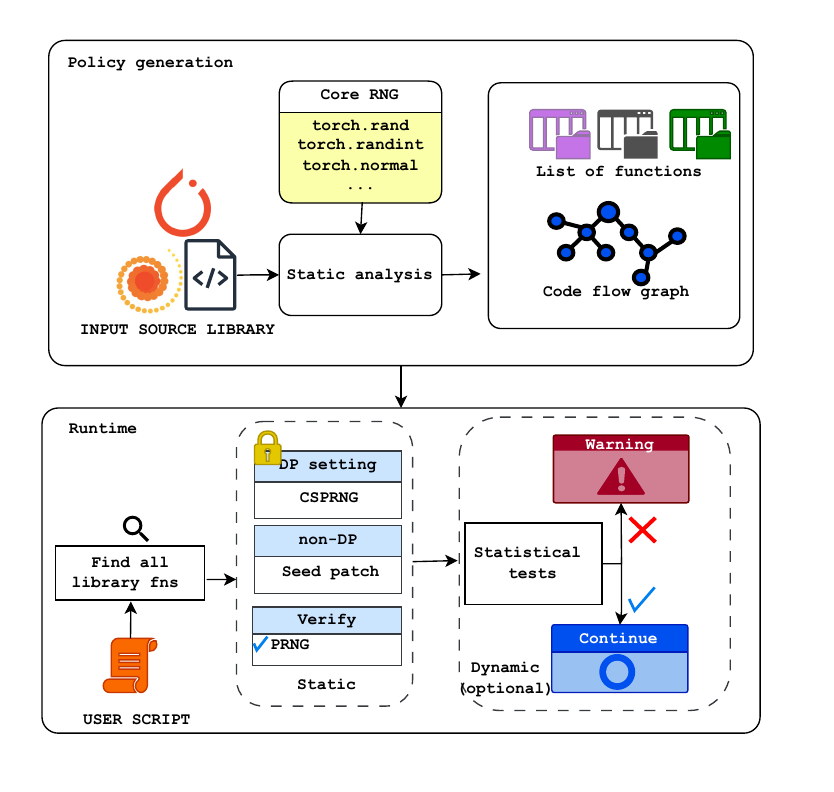}
    \caption{\sys works in two phases - policy generation and runtime secure enforcement. Given a set of core RNG functions, \sys identifies all functions in the target framework that depend on the core RNG functions using static analysis. At runtime, \sys can operate in static and dynamic modes by enforces policies derived from our ML ecosystem survey and running statistical tests respectively.
    %on the generated noise from the identified function, raising a warning if a deviation is detected.
    }
    \label{fig:sys}
   \vspace{-.5cm} 
\end{figure}
\subsection{\sys Policy Generation}
\label{sec:rngguardcore}

%\todo{
The first phase of \sys consists of policy generation for all randomness sources in the framework, as depicted in Figure~\ref{fig:sys}.
\sys uses static analysis to find all functions that directly or indirectly rely on core PRNG functions, which in turn are identified through  
%We identify the core RNG functions using 
our ecosystem study (as summarized in Tables~\ref{table:MLpipelineRand} and~\ref{table:impl}), guided by previous knowledge of the framework and name-based heuristics. Once identified, these functions can be associated with the distributions they generate (e.g. uniform or Gaussian). We use CodeQL~\cite{codeql}, a semantic code query engine, to scan the source code of the target framework. We start with the set of core PRNG functions followed by CodeQL's transitive analysis to extract all the higher-level functions that depend on these primitive functions. This process allows us to determine what components require replacement or auditing at runtime. For each call to a core RNG function, RNGGuard can check whether the output distribution matches the expected distribution identified in the survey by checking if it calls the correct core RNG functions with the correct distribution parameters (if known) and if it uses a safely-seeded PRNG or a CSPRNG. % and optionally run it through the distribution tests.
As a proof-of-concept, we apply \sys over the PyTorch ecosystem (including Opacus). We note that
\sys can be applied to other ML frameworks using the same steps.
%We identify a set of core random functions by inspecting the official PyTorch API reference \cite{PyTorchCitation} - 
The set of core random functions identified for PyTorch include 
 \texttt{torch.rand, torch.randint, torch.randperm, torch.normal, torch.randn} and their in-place counterparts, \texttt{torch.Tensor.normal\_ and torch.Tensor.uniform\_}.

\subsection{\sys Runtime Operating Modes}
In the second phase, \sys implements runtime secure enforcement as presented in Figure~\ref{fig:sys}. \sys relies on Python's dynamic nature to enforce secure execution and add statistical testing to the core randomness functions identified in static analysis. Python allows for the replacement of entire functions in memory without having to modify source code at runtime. \sys leverages this property together with Python's module caching system to enable easy auditing and code instrumentation without having to rebuild libraries. 
%\sys also takes advantage of Python's module caching system. 
When a module is first imported into Python, it is loaded into a cached object and all subsequent imports will use that cached module instead of re-importing a library. 
As a result, integrating \sys is as simple as importing it first in any script before any target library modules are loaded.
%When using \sys, it must be imported first, before importing any PyTorch module; this allows for the patched PyTorch library to be cached. Any libraries depend on PyTorch (say, Opacus) will use the cached reference. To ensure correct patching, \sys must be imported before any PyTorch modules are loaded. 
%
% \todo{let's avoid using "patch". we can say policy or secure execution enforcement}
\sys supports two operating modes: 

\mlparagraph{Static.} In this mode, \sys ensures secure randomness by enforcing policies derived from our requirement analysis and ecosystem study. For example, \sys applied on PyTorch will automatically inject a generator seeded from 
\texttt{/dev/urandom} if there is no generator passed into a function call and replaces
PyTorch's RNG functions with a CSPRNG for DP. 
%It also automatically injects a generator seeded from \texttt{dev/urandom} if there is no generator passed into a function call. 
Additionally, \sys will also enforce secure noise generation in Opacus via its API. 

\mlparagraph{Dynamic.} In this mode, \sys enables runtime statistical tests (described in Section~\ref{subsec:stataudit}) for both discrete and continuous distributions. While the runtime tests introduce a higher runtime overhead, they provide more flexibility and can be extended to scenarios where access to source code is limited. To improve the runtime overhead of \sys's dynamic mode, we introduce system optimizations as presented later. 
\sys integrates the KS test~\cite{KStest} and Pearson's $\chi^2$ test~\cite{PearsonChi} for continuous and discrete distributions respectively.
%(as in the case of \texttt{torch.randperm}) from the \textit{scipy} \cite{scipy} library as these tests are adopted in other ML frameworks (table \ref{table:frameworktests}).
The Pearson $\chi^2$ test requires a minimum count of 5 observations per bin and minimum sample size of 13 to be valid, while the KS test on SciPy~\cite{scipy} requires a sample size of 20. To this end, we use a sample size of 100 for both statistical tests and map all continuous observations to $\mathcal{U}(0,1)$ or $\mathcal{N}(0,1)$ (as described in Section~\ref{sec:randomnessfromdist}). Detecting a deviation from the expected distribution ($p < 0.01$) raises a warning. 
%Evaluating other statistical tests (e.g., Anderson–Darling \cite{adtest}, Cram\'{e}r–von Mises \cite{cramer} etc.) and exploring their applicability and effectiveness in auditing PRNG behavior is a direction for future work. 

\section{Evaluation}

\subsection{Experimental Setting}
\label{sec:evalsetting}
We use Python 3.12 with PyTorch 2.8 compiled with CUDA 12.6 and Opacus 1.5.4. 
Due to compatibility issues, we manually build torchcsprng 0.3.a from source.
%We follow the training configuration provided in the official Opacus documentation~\cite{opacus-tutorial}. 
For DP-SGD training, we use a clipping norm $1.2$, target privacy budget of $\epsilon = 50.0$, and $\delta = 10^{-5}$ to train a ResNet20 model \cite{resnet} on the CIFAR10 dataset~\cite{cifar}.
%with a batch size of 128, and run experiments on an
Our experiments are reported on a NVIDIA A30 GPU. We use statistical tests based on implementations from SciPy. To assess the practical overhead imposed by \sys in realistic ML workflows, we evaluate runtime for model training in DP and non-DP settings. We then present two optimizations in section \ref{para:results} to address the overhead of \sys dynamic operating mode and present results for the DP setting. 

\begin{table}[ht]
\caption{Number of violations reported by Delta-Siege over ten runs auditing Opacus Gaussian noise for
%with Delta-Siege for each given privacy guarantee. We report the number of violations in four different settings: 
CSPRNG/non-CSPRNG with and without FP attack mitigation.
%The number of violations is reported in terms of detected violations over ten independent runs for that particular privacy guarantee.
}
\label{tab:DPAuditTools}
% \begin{minipage}{\columnwidth}
\resizebox{.5\textwidth}{!} {
    \begin{tabular}{ccccc}
        \toprule
         & \multicolumn{4}{c}{\textbf{Number of violations}} \\
         \cmidrule{2-5}
        \textbf{$(\epsilon, \delta)$} & \makecell{\textbf{CSPRNG} \\ \textbf{FP} } & \makecell{\textbf{non-CSPRNG} \\ \textbf{FP}} & \makecell{\textbf{CSPRNG} \\ \textbf{no FP}} & \makecell{\textbf{non-CSPRNG} \\ \textbf{no FP}}  \\ \midrule
        (0.1, $1\times 10^{-6}$)  &3& 1 & 9 & 10 \\ \midrule
        (0.1, 0.001)  & 4 & 2 & 10 & 10\\ \midrule
        (0.1, 0.1)  & 7& 8 & 10 & 10\\ \midrule
        (1, $1\times 10^{-6}$) & 10 & 10 & 10 & 10 \\ \midrule
        (1, 0.001) & 10 & 10 & 10 & 10\\  \midrule
        (1, 0.1) & 2 & 1 & 10 & 10\\ 
        
        \bottomrule
           
    \end{tabular}
    }
%    \end{minipage}
\end{table}

\subsection{DP Auditing Evaluation}
%\todo{
We start by evaluating the effectiveness of DP auditing tools in detecting CSPRNG security issues.
%, and evaluate its effectiveness to detect CSPRNG security issues. 
We run DeltaSiege~\cite{DeltaSiege} on the Opacus Gaussian Mechanism, with and without CSPRNG, for six different values of $(\epsilon, \delta)$ as described in Table~\ref{tab:DPAuditTools}. We further separate non-CSPRNG and CSPRNG implementations with and without floating-point (FP) attack (~\cite{mironov, holohan2024securing}) mitigation. 
%FP attacks~\cite{mironov, holohan2024securing} is a class of attacks that exploit the limited precision of 32 and 64-bit floats used to represent DP noise. FP rounding and finite sampling introduce further loss in precision in the noise, making some outputs distinguishable. 
The root cause analysis presented by DeltaSiege~\cite{DeltaSiege} only focuses on FP attacks and attributes privacy violations to FP arithmetic~\cite{holohan2024securing} and noise sampling vulnerabilities~\cite{holohan2021securerandomsamplingdifferential}. %Opacus~\cite{opacus} implements only one noise sampling  mitigation~\cite{holohan2021securerandomsamplingdifferential}. 
% As previously discussed in section \ref{para:dpaudit}, the determination of privacy violation is empirical and may vary across runs. Therefore, we report the number of violations detected from ten independent runs for each $(\epsilon, \delta)$ in Table \ref{tab:DPAuditTools}. 
%
%Our evaluation demonstrates that DeltaSiege is not able to detect vulnerabilities due to non-secure PRNG usage.
As our results in Table~\ref{tab:DPAuditTools} demonstrate, the number of violations for FP and no FP mitigation cases remain roughly the same whether or not a CSPRNG is applied, indicating that  
%with and without the CSPRNG remains roughly the same in both cases whether the FP mitigation is applied or not. In other words, 
DeltaSiege cannot detect additional violations due to the underlying source of randomness.
%}

% We start by evaluating a DP auditing tool against security vulnerabilities studied in this paper.

% We report the results in Table~\ref{tab:DPAuditTools}. As previously discussed in section \ref{para:dpaudit}, the determination of privacy violation is empirical and may vary across runs. Therefore, we report the number of violations detected from ten independent runs in Table \ref{tab:DPAuditTools} . 
% While Delta-Siege detected privacy violations in all tested settings, it had the same number of violations in some cases regardless of whether the mechanism used a CSPRNG. As a result, it could not flag the use of insecure randomness in these cases. 

\subsection{\sys Evaluation}

\mlparagraph{Core RNG Microbenchmarks.}
We evaluate the runtime of core RNG functions identified in section \ref{sec:rngguardcore} by measuring the time required to generate and test 1000 samples on the CPU in Table~\ref{tab:compres}. For \texttt{torch.randperm}, we measure the time taken to shuffle a 1D tensor of shape $(1000, )$.
%filled with ones. 
%We report average runtime per call over 1000 function calls in Table~\ref{tab:compres}. 
\sys's static mode introduces moderate overhead ($46-112\%$) due to the creation and injection of new generators.
%for every function call. 
The dynamic mode incurs a higher runtime cost, often exceeding 1 ms per function call due to \sys testing all instances of generated noise. The in-place noise generation functions, \texttt{Tensor.normal\_} and \texttt{Tensor.uniform\_} have the highest overhead in the dynamic mode.
%, upto 12400\%. 

\begin{table}[t]
\caption{Runtime for PyTorch Core RNG functions (average over 1000 calls) for the baseline and \sys modes.}
\label{tab:compres}
\resizebox{.5\textwidth}{!} {
    \centering
    \begin{tabular}{ccccc}
    \toprule
        \textbf{Component} & \makecell{\textbf{Runtime} \\ \textbf{(ms/call)}} & \makecell{\textbf{Static} \\ \textbf{(ms/call)}} & \makecell{\textbf{Dynamic} \\ \textbf{(ms/call)}}  \\
        \midrule
        \texttt{rand} &0.015 & 0.022  &  1.266 \\ \midrule
        \texttt{randn} & 0.016 & 0.024  &1.581 \\ \midrule
         \texttt{randint} & 0.025 & 0.032  & 1.014\\ \midrule
         \texttt{randperm} & 0.010    &  0.017   & 0.506 \\ \midrule
         \texttt{normal} & 0.016  & 0.025 & 1.455 \\ \midrule

          \texttt{Tensor.uniform\_} & 0.008    & 0.016      &0.864    \\ \midrule
        \texttt{Tensor.normal\_} & 0.008   &0.017    &1.000     \\ \midrule
    \end{tabular}
    }
\end{table}

\begin{figure}[t]

\centering
\includegraphics[width=\linewidth]{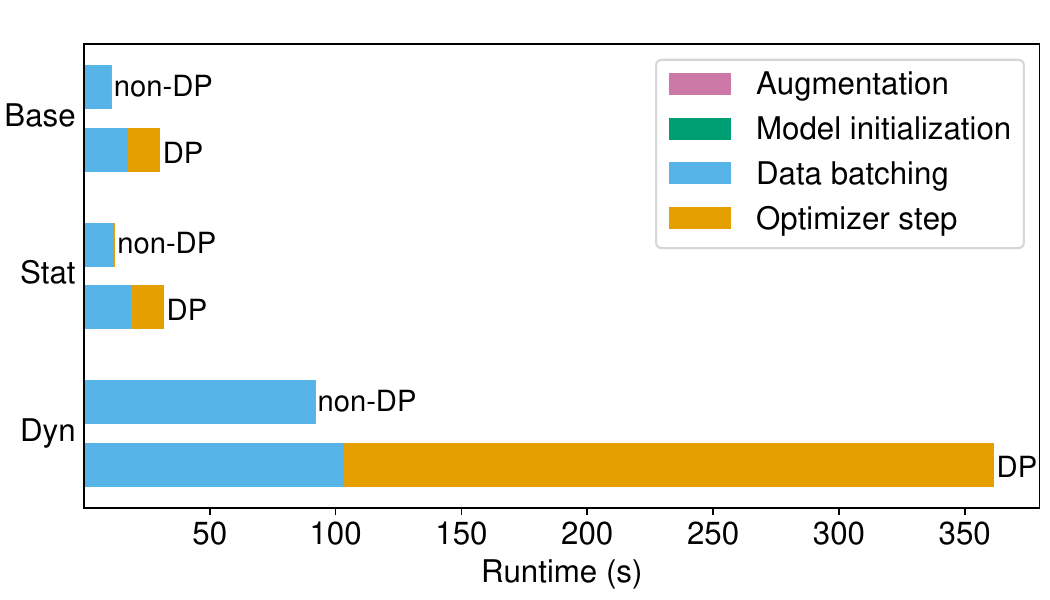}%

\caption{Runtime of \sys for one epoch  training in non-DP and DP settings showing baseline (Base) and \sys static (Stat) and with dynamic (Dyn) modes.}
\label{fig_compare}

\end{figure}

\mlparagraph{End-to-end Training Benchmarks.}
Figure~\ref{fig_compare} depicts the overhead of \sys for end-to-end model training (one epoch) comparing baseline and \sys static and dynamic modes for non-DP and DP settings.
%We evaluate \sys in end-to-end model training for CIFAR10 on ResNet20, and compare total training time for one epoch among the baseline (unmodified), \sys static mode, and \sys dynamic mode in Figure~\ref{fig_compare}.
% Figures \ref{fig_nondp} and \ref{fig_dp}. 
\label{para:results}

%We consider the following two settings for the end-to-end training:
\paragraph{Non-DP Setting} The baseline runs in $15 \ s$, while the static and dynamic modes take $22 \ s$ ($47\%$ overhead) and $132 \ s$ respectively. 
%significant runtime slowdown of 780\%. 
In all three cases, the data batching process consumes up to $98\%$ of the runtime followed by the SGD optimizer step, data augmentation, and model initialization. 
\paragraph{DP Setting} For the static (dynamic) operating mode, data batching takes $80\%$ ($70\%$) of training time followed by DP-SGD optimizer step taking up to $19\%$ ($29\%$). Since model initialization and data augmentation are one-time steps, they \emph{contribute minimally} to the training cost. The overall runtime overhead of static and dynamic modes over the baseline are $16.5\%$ and $650\%$ respectively, highlighting the higher cost of runtime auditing. To address this, we introduce optimizations for \sys dynamic mode as described next.
%without sacrificing statistical coverage or security.

\subsection{\sys Optimizations}
\begin{figure}[t]
    \centering
    \includegraphics[width=\linewidth]{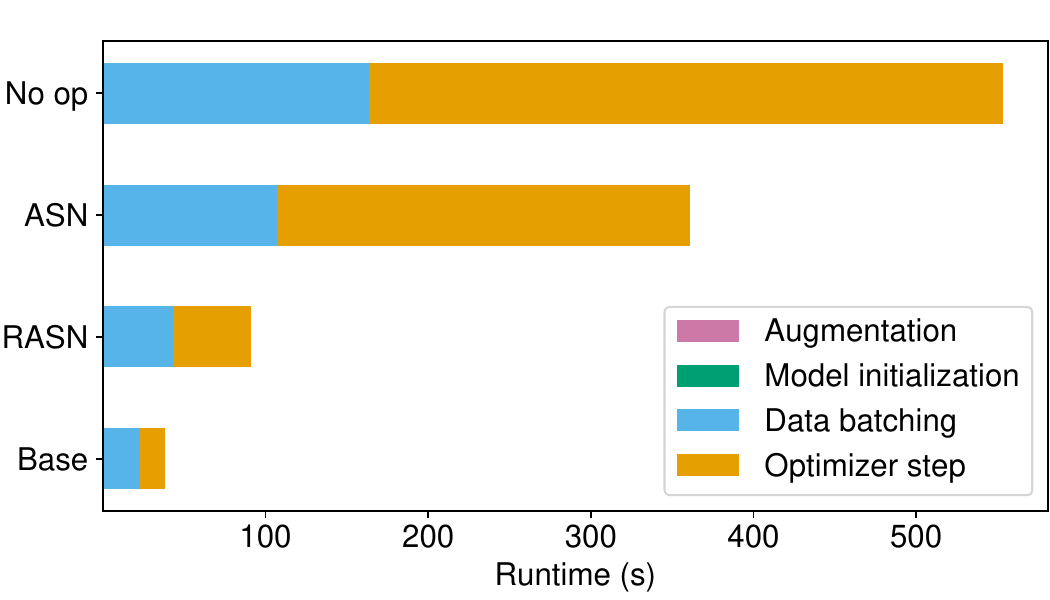}
    \caption{Runtime of \sys for one epoch of DP training showing baseline (Base), No optimization (No op), as well as ASN and RASN optimizations.
    %after optimizations. We present runtimes for DP-ML training for the baseline (Base) and when \sys has no optimizations (No op). Then we present the runtimes for both ASYNC (ASN) and Randomized ASYNC (RASN) modes. Note that \sys is in Dynamic mode here in scenarios where it's enabled.
    }
    \label{fig:newdpopt}
\end{figure}

In this section we present our iterative optimizations over dynamic mode and benchmark results in Figure~\ref{fig:newdpopt}. 
% Our optimizations aim to perform tests for randomness on the generated noise prior to use, to ensure they conform to the desired random distribution (Gaussian or Uniform). 
We present runtime evaluations for
%As the core RNG functions are used in the DataLoader and DP-SGD optimizer, we present runtimes for 
data transformations, model initialization, data batching, and DP-SGD optimizer step.
%, and the overall DP training time over the entire dataset. 

\mlparagraph{Baseline.} This setup performs DP training without \sys, with the GPU handling most computations during model training.
%the model on incoming batches of data  the CPU remains occupied with minimal to no workloads during training.
In this setting, data shuffling in the DataLoader takes around $23 \ s$ whereas the DP-SGD step takes $15 \ s$, contributing to the majority of the DP training runtime of $55 \ s$.

\mlparagraph{No Optimization.} This multi-threaded implementation launches a CPU thread for statistical testing, i.e., the \emph{auditor}.
%, synchronizing GPU-based training and CPU-based statistical testing. 
Once the number of random values generated by the CSPRNG on the GPU exceeds the test sample size, they are copied to CPU memory and placed in a work queue. The auditor consumes random numbers from the work queue and runs the statistical test during which the main process
%places the random numbers on the queue, 
halts execution.
%and waits for the test to be completed by the auditor. 
After computing the statistic and p-value, the auditor inserts the results into a result queue for retrieval.
The runtime for data batching, DP-SGD, and overall training in this setting are $163 \ s$, $390 \ s$, and $571 \ s$ respectively.

\mlparagraph{ASN mode.} The Asynchronous (ASN) mode operates similar to above with the main process accumulating and submitting random numbers to the work queue. However, the main process does not halt execution and proceeds with training while the auditor consumes numbers. 
%Consequently, the main process does not verify the computed test statistic to determine the conformity of the random numbers to the expected distribution; the auditor makes this determination. 
By allowing the training and statistical testing to proceed concurrently, we observe runtime speedup where the data shuffling, DP-SGD, and overall training times drop to $107 \ s$, $254 \ s$, and $374 \ s$ respectively.

\mlparagraph{\textsc{RASN} mode.} Our final optimization in the Randomized Asynchronous (RASN) mode consists of only testing a subset of the random numbers generated. Specifically, we audit every tenth random sequence generated by the CSPRNG. To achieve this, the main process maintains a counter for the generated sequences by the CSPRNG. When the counter reaches a multiple of ten, the current sequence is added to a local buffer
%we select that sequence for auditing and add it to a local buffer 
until its length meets the test sample size for consumption by the auditor.
%The sequence is then sent to the auditor via the work queue for testing. 
RASN further reduces the runtime to $43 \ s$, $48 \ s$, and $108 \ s$ for data shuffling, DP-SGD, and overall DP training.

\subsection{\sys Soundness Evaluations}
\mlparagraph{Function Coverage.}
Once the core RNG primitives are protected (e.g., via safe seeding or replacement with a CSPRNG), all downstream functions inherit the same guarantees.
%\Paragraph{Attack.} \todo{
To demonstrate \sys's effectiveness, we implement an attack which modifies the Kaiming Normal weight initialization function (\texttt{torch.nn.init.kaiming\_normal\_}) to return uniformly distributed weights. RNGGuard can detect this replacement as the attacker’s function does not call the right RNG primitive and it fails the KS test in the dynamic mode. After \sys enforces its security policies, the weight initialization function passes the normal distribution test again. To additionally verify functionality, we perform 25 epochs of end-to-end training with the setup described in Section~\ref{sec:evalsetting}. The model trained with the original initialization (no attack) achieves an accuracy of $21.09\%$ while the model trained with the secured initialization function achieves $21.37\%$, indicating that the \sys enforcement is correct and preserves functionality.
%}
% \zahra{we don't have a related work section?!}
\section{Discussion}
%\todo{
Our results show that randomness auditing and policy enforcement is a promising direction to guarantee secure randomness in ML systems. The primary cost of dynamic auditing through statistical tests arises from the synchronization between GPU training and CPU testing. Decoupling these components and allowing them to run concurrently substantially improves the throughput, with additional speedup achieved in the RASN mode. Future work can further reduce this overhead by accelerating tests on the GPU.

Beyond performance, \sys provides sound enforcement by securing core RNG primitives and automatically restoring tampered functions while preserving their functionality. While our evaluations focused on PyTorch, \sys can generalize across frameworks. To test generalizability, we also integrate \sys into FedML-AI~\cite{fedmlai} which uses PyTorch RNG primitives but does not use CSPRNGs. \sys enforces its policies on the DP noise addition mechanisms to use CSPRNG and the resulting secure functions pass all existing unit and statistical tests.  

%We note that while \sys's dynamic mode introduces overhead, statistical tests can be run asynchronously so as to not affect overall throughput, and future work can reduce this overhead by accelerating tests on the GPU. 
Finally, we note that while identifying core RNG functions currently requires some manual analysis, future work can extend \sys with techniques to automate this step. 
%}
\section{Related Work}

%generic attacks on randomness, 1 on ML auditing generic, randomness tests, attacks on ML randomness, 
\mlparagraph{Attacks on Randomness: } Cryptographic literature has extensively studied attacks on PRNGs. These include state extension and seed recovery attacks\cite{mt19937attack} where an attacker can recover the internal state by observing outputs from legacy PRNGs such as MT19937 \cite{MT19937} PRNG. Other attacks exploit poor seeding practices to decrypt network traffic\cite{10.1145/3243734.3243756} and recover private keys \cite{systemtimeatk, 10.1145/3243734.3243756}. Additionally, prior work~\cite{lowentropy, kaplanlinuxprng} has exploited low entropy at system boot to recover several cryptographic keys~\cite{lowentropy} and launch denial-of-service attack~ \cite{kaplanlinuxprng} using OS-derived randomness.

\mlparagraph{Randomness Tests: } Randomness testing frameworks evaluate whether a sequence of numbers conforms to certain statistical properties. Common statistical tests~\cite{KStest, PearsonChi, adtest, cramer} check if a sample distribution matches a reference distribution through hypothesis testing. The NIST SP800-22 test suite~\cite{NIST800} includes a comprehensive set of statistical tests designed to evaluate uniformity, independence, and entropy of random sequences. 
%They also include guidelines and recommendations for secure randomness generation. 
However, 
%as demonstrated in some attacks~\cite{dahiyarandomness}, 
a PRNG that passes these test suites is not necessarily always secure in practice, as both secure and insecure RNGs can sometimes pass the same tests~\cite{dahiyarandomness}.

\mlparagraph{Static Analysis for Insecure Randomness: } Tools like CodeQL~\cite{codeql}, FindSecBugs ~\cite{findsecbugs}, and TaintCrypt~\cite{taintcrypt} explore automated detection of insecure randomness and RNG misuse via static analysis.
CodeQL~\cite{codeql} and FindSecBugs~\cite{findsecbugs} flag the usage of insecure randomness (e.g. Python’s random API or C++’s std::mt19937 API etc.) in security-sensitive contexts by name-based heuristics~\cite{codeql} and pattern detection~\cite{findsecbugs}. Other tools such as TaintCrypt~\cite{taintcrypt} use taint analysis to identify functions that depend on randomness and detect potential vulnerabilities arising due to RNG API misuse or insecure random APIs. 
Our work differs from prior approaches by providing a unified system for both static enforcement and dynamic auditing of randomness in ML frameworks. Unlike prior DP auditors, \sys can detect the use of insecure PRNGs and enforce cryptographically secure randomness.

% \zahra{maybe add another paragraph on ml security/privacy in general and attacks/mitigations}
%\todo{
\mlparagraph{ML Security and Privacy Attacks: } A large body of work has shown that ML systems are vulnerable to a range of security and privacy threats, including poisoning~\cite{jha2023labelpoisoningneed, shafahi2018poisonfrogstargetedcleanlabel}, %backdooring~\cite{rance2022augmentationbackdoors, badnets}, 
model evasion~\cite{modelevasion2, modelevasion1}, and privacy~\cite{membershipinference, lira}. Prior works have proposed mitigation techniques such as outlier removal~\cite{datapoisoningmit} against poisoning,
%, trigger detection and removal for backdoors~\cite{BackdoorMit1}, 
adversarial training for evasion~\cite{RSmoothingOG}, and privacy-preserving training mechanisms like DP~\cite{DPSGD} to reduce membership leakage.
Cryptographic techniques such as signatures~\cite{gan2025sentry} or secure multiparty computation~\cite{almashaqbeh2023anofel} and their acceleration on GPUs~\cite{gan2025sentry,gan2025cuot,spoczynski2025scalable} have also been explored to provide security and privacy in ML systems. 
%}

\mlparagraph{DP and DPSGD Auditing Tools: } DP auditing tools \cite{DeltaSiege, DPAuditorium} empirically evaluate privacy guarantees by generating neighboring dataset pairs as inputs to the mechanism and testing  whether an adversary can distinguish between the outputs. 
%However, auditing mechanisms like DPSGD can be hard \cite{dpsgdaudit, dpauditsurveypaper}; 
Some DPSGD auditors~\cite{murakonda2020mlprivacymeteraiding, dpsgdaudit} test DPSGD implementations by comparing noisy gradients across neighboring training datasets, while others leverage existing attacks like membership inference attacks~\cite{membershipinference} to estimate privacy risk. These tools generally do not audit the underlying PRNG used to generate DP noise, which may leave ML pipelines vulnerable if weak or insecure randomness is deployed. 

\section{Conclusion}
In this work we examined the role of randomness in ML systems and conducted an ecosystem study on practices and implementations in major ML frameworks. Our study revealed various issues in existing systems including poor design choices or insecure implementations as it relates to PRNG seed and algorithms. Furthermore, we proposed an attack taxonomy based on practical settings, and derived security policies for randomness in ML systems. We proposed \sys as a system to enforce security policies over randomness. \sys performs a static analysis to identify instancies of random functions, and enforces secure execution at runtime. %We believe insights and products from this work can contribute towards the security of ML systems. 

%%%%% TODO
%\ifCLASSOPTIONcompsoc
%  % The Computer Society usually uses the plural form
%\else
%  % regular IEEE prefers the singular form
%\fi

% \itodo{DONE} \zahra{@software for citation may not be supported, you can just use misc instead}
% \itodo{DONE}  \zahra{citation \cite{sfc} is overflowing, use a url line breaker} 
\bibliographystyle{IEEEtran}
%argument is your BibTeX string definitions and bibliography database(s)
% \itodo{capitalize the citations when needed for e.g. TSL instead of Tsl}
\bibliography{IEEEabrv,papers}
% \appendix
% %% fmt todo: figure the appendix ieeetran
% \section{a}
% \tentative{Add the LLM decoding step evaluations}

% \tentative{Add Generalizability results here}

\end{document}